\documentclass[twocolumns]{aa}

\def\kms{$\rm km\;s^{-1}$}
\def\dg{^\circ}
\def\ha{H$\alpha$}
\def\h2{H$_{2}$}
\def\hi{H~{\scriptsize I}}

\def\ng{[N~{\scriptsize II}]$\,\lambda6583.41$}
\def\msun{M$_{\odot}$}
\def\lbsun{L$_{B,\odot}$}

\begin{document}

\thesaurus{
	      03(11.09.1: NGC 4672; 
		 11.11.1;           
		 11.19.2;           
                 11.06.1;           
		 11.19.6)}          

\title{NGC~4672: a new case of an early-type disk
galaxy with an orthogonally decoupled core\thanks{Based on
observations carried out at ESO, La Silla 
(Chile).}$^{\bf,}$\thanks{Tables 3 to 6 are only available in 
electronic form at the CDS via anonymous ftp to cdsarc.u-strasbg.fr 
(130.79.128.5) or via http://cdsweb.u-strasbg.fr/Abstract.html.}}

\subtitle{}

\author{
     M.~Sarzi             \inst{1},
     E.M.~Corsini         \inst{1}, 
     A.~Pizzella          \inst{1},
     J.C. Vega Beltr\'an  \inst{2},
     M.~Cappellari        \inst{3}, 
     J.G.~Funes~S.J.      \inst{3,4},  
     and F.~Bertola       \inst{3}}
	
\offprints{sarzi@pd.astro.it}

\institute{
Osservatorio Astrofisico di Asiago, Dipartimento di Astronomia,
Universit\`a di Padova, 
via dell'Osservatorio~8, I-36012 Asiago, Italy \and
Instituto Astrof\'\i sico de Canarias, Calle Via Lactea s/n,
E-38200 La Laguna, Spain \and
Dipartimento di Astronomia, Universit\`a di Padova, 
vicolo dell'Osservatorio~5, I-35122 Padova, Italy \and
Vatican Observatory, University of Arizona, 
Tucson, AZ 85721, USA}

\date{Received 15 February 2000; accepted...................}

\authorrunning{Sarzi et al.} 
\titlerunning{NGC 4672}

\maketitle

\begin{abstract}

We report the case of the early-type disk gala\-xy NGC 4672 as a new
example of a galaxy characterized by the orthogonal geometrical
decoupling between bulge and disk. The morphological features of this
galaxy are discussed as well as the velocity curves and velocity
dispersion profiles of stars and ionized gas along both its major and
minor axis. We conclude that NGC 4672 has structural ({\it i.e.\/} a
bulge elongated perpendicularly to the disk) and kinematical ({\it
i.e.\/} a stellar core rotating perpendicularly to the disk) properties
similar to those of the Sa NGC 4698. The presence of the isolated
core suggests that the disk component is the end result of the
acquisition of external material in polar orbits around a pre-existing
oblate spheroid as in the case of the ring component of AM 2020-504,
the prototype of polar ring ellipticals.


\end{abstract}

\section{Introduction}
\label{sec:n4672_introduction}

Most of the kinematically decoupled structures observed in ga\-laxies
have been considered to have an external origin, and to represent
the end result of the acquisition of material coming from outside the
boundaries of an already formed galaxy via infall, accretion or
merging. The angular momenta of the acquired and pre-existing material
are generally misaligned with respect to each other. In fact the
angular momentum of the acquired material depends on the impact
parameters of the capture process and on the geometry of the potential
of the host galaxy. This is the case of the elliptical galaxies with a
dust lane along their minor axis (see Bertola 1987 for a review) and
of the polar-ring galaxies (see the photographic atlas of Whitmore et
al. 1990), where the acquired gas (and its associated stars) is
rotating around an axis orthogonal to the minor axis of the main
stellar body.  Moreover this is also the case of the Sa galaxy
\object{NGC 4698}, where the presence of an orthogonal decoupling
between bulge and disk has been recently discussed by Bertola et
al. (1999).

The mass of the orthogonally rotating material is negligible ($M_{\rm
gas}\la\,10^9$ \msun) with respect to the total dynamical mass of the
host galaxy in the minor-axis dust-lane elliptical galaxies (Sage \&
Galletta 1993). The mass of the acquired gas ($M_{\rm
gas}\approx\,10^{10}$ \msun) is comparable or even greater than the
mass of the pre-existing galaxy in the polar-ring galaxies, like the
elliptical \object{AM 2020-504} (Arnaboldi et al. 1993a), the S0
\object{NGC 4650A} (Sackett et al. 1994) and the spiral \object{NGC
660} (van Driel et al. 1995).  In NGC 4650A, the polar structure is
massive enough to lead to the formation of a spiral pattern, as
pointed out by Arnaboldi et al. (1997). This is a strong indication
that in this object the acquired material settled onto a disk rather
than forming a ring. Besides, there is a number of similarities
between the global properties ({\it e.g.\/}, total luminosity,
integral colors, gas-to-dust ratio and star formation rate) of the
polar-rings around S0's, and spirals and the properties of the
disks of late-type spiral galaxies (see Reshetnikov \& Sotnikova 1997
and references therein).
 
The case of the Sa galaxy NGC 4698 (Bertola et al. 1999) could
represent the first example of an orthogonal acquisition involving an
even larger amount of external material, which produced an intrinsic
change of the galaxy morphology. In fact the geometric and kinematical
orthogonal decoupling between the bulge and disk of this early-type
spiral galaxy has been interpreted as the end product of the accretion
of material which built up the disk component around a previously
formed bare spheroid.

In the framework of massive acquisition processes, we pre\-sent
\object{NGC 4672} as a new case of a disk
galaxy characterized by a geometric and kinematical orthogonal
decoupling between its bulge and disk. This paper is organized as it
follows. In Sect. \ref{sec:n4672_observations} we present our
spectroscopic observations of NGC 4672. The morphology of this object
and its classification are discussed in Sect.
\ref{sec:n4672_morphology}.  The kinematical results for both the
gaseous and the stellar components are described in Sect.
\ref{sec:n4672_kinematics} and we move
towards our conclusions in Sect. \ref{sec:n4672_conclusions}.

\section{Observations and data reduction}
\label{sec:n4672_observations}

The spectroscopic observations of NGC 4672 were carried out at the
European Southern Observatory in La Silla during three different runs.
We observed the galaxy at the Danish 1.54-m telescope on July 12, 1998
and at the ESO 1.52-m telescope on August 6, 1998 and on June 9-10,
1999 respectively.

The Danish 1.54-m telescope was equipped with the Danish Faint Object
Spectrograph and Camera (DFOSC). We used the grism No.~7 with 600
grooves $\rm mm^{-1}$ and the Loral/Lesser C1W7 CCD, which has
$2052\;\times\;2052$ pixels of $15\;\times\;15$ $\rm
\mu m^2$. 
We took six separate spectra of 30 minutes each along the minor axis
of the galaxy ($\rm P.A.=134\dg$) for a total exposure time of 3
hours. To position the slit on the galaxy center an acquisition image
of 300 s was obtained in the $R$ band using the ESO filter No.~452.

The Cassegrain Boller \& Chivens spectrograph was mo\-un\-ted at the
ESO 1.52-m telescope in combination with the grating No.~33 with 1200
grooves $\rm mm^{-1}$ and the Loral/Lesser CCD No.~39 with
$2048\;\times\;2048$ pixels of $15\;\times\;15$ $\rm \mu m^2$.  We
obtained a single 60-minutes spectrum and 4 separate spectra of 60
minutes along the galaxy major axis ($\rm P.A.=46\dg$) in August 1998
and June 1999, respectively.

Every night a number of spectra of late-G and early-K giant stars were
obtained to be used as template in measuring the stellar
kinematics. Comparison spectra were taken before every object
exposure. Further details of the instrumental set-up in the three
observing runs are given in Table~\ref{tab:n4672_setup}.

\begin{table*}[ht!]
\caption{Instrumental set-up of spectroscopic observations}
\begin{center}
\begin{tabular}{lccc}
\hline
\noalign{\smallskip} 
\multicolumn{1}{l}{Parameter} & 
\multicolumn{1}{c}{$\rm P.A.=134\dg$} & 
\multicolumn{2}{c}{$\rm P.A.=46\dg$} \\
\noalign{\smallskip} 
\hline
\noalign{\smallskip} 
Date                   & Jul 12, 1998  & Aug 06, 1998 & 
  Jun 09-10, 1999 \\ 
Instrument             & Danish 1.54-m$+$DFOSC        & 
  \multicolumn{2}{c}{ESO 1.52-m$+$B\&C} \\
Scale                  & $\rm 0\farcs39\;pixel^{-1}$ & 
  \multicolumn{2}{c}{$\rm 0\farcs82\;pixel^{-1}$} \\
Reciprocal dispersion  & $\rm 1.47 \AA\;pixel^{-1}$    & 
  \multicolumn{2}{c}{$\rm 0.98 \AA\;pixel^{-1}$} \\
Slit width             & $1\farcs5$  & $2\farcs4$ &
  $2\farcs2$ \\
Wavelength range       & 3847 -- 6850 \AA\        & 4858 -- 6872 \AA\ & 
   4813 -- 6829 \AA\  \\
Comparison-lines FWHM  & $5.58\pm0.02$ \AA\       & $2.77\pm0.08$ \AA\ & 
  $2.86\pm0.04$\\ 
Instrumental $\sigma$ at \ha\    & $108$ \kms\    & $54$ \kms\ & 
  $55$ \kms\   \\
Exposure time          & $6\times30$ min          &  $1\times60$ min &
  $4\times60$ min \\
Seeing FWHM            & $1\farcs8$ -- $2\farcs2$ & $1\farcs5$ -- $2\farcs0$ &
  $1\farcs5$ -- $2\farcs5$\\
\noalign{\smallskip} 
\hline
\end{tabular}
\end{center}
\label{tab:n4672_setup}
\end{table*}

Using standard ESO-MIDAS\footnote{MIDAS is developed and maintained by
the European Southern Observatory} routines all the spectra were bias
subtracted, flatfield corrected, cleaned from cosmic rays and
calibrated. The spectra taken along the same axis in the same run were
co-added using the center of the stellar continuum as reference. The
contribution of the sky was determined from the edges of the resulting
spectra and then subtracted.

The stellar kinematics was measured from the absorption lines present
in the spectra using the Fourier Correlation Quotient Method (Bender
1990) as applied by Bertola et al. (1996). The values obtained for the
stellar kinematics along the major and minor axis are reported in
Tables 3 and 4, respectively.
The ionized-gas kinematics was measured from the \ha\ and \ng\
emission lines by means of the MIDAS package ALICE as done by Corsini
et al. (1999). The gas velocity and velocity dispersion are the mean
values obtained from the two emission lines. No error is given when
only one velocity or velocity dispersion measurement is available. The
kinematical data derived for the gaseous component along the major and
minor axis are given in Tables 5 and 6, respectively.  

The $R-$band acquisition image (Fig.~\ref{fig:n4672_cont}) was bias
subtracted and fla\-tfield corrected.  No attempt was made to perform
an absolute calibration since the image was obtained under
non-photome\-tric condition.

\section{The morphology of NGC 4672}
\label{sec:n4672_morphology}

NGC 4672 (ESO~322-G73) is a highly-inclined early-type disk galaxy
classified  as Sa(s) pec sp by de Vaucouleurs et al. (1991). An
overview of  the properties of NGC 4672 is given in Table
\ref{tab:n4672_properties}.

\begin{figure}[ht]
\vspace*{11cm}
\includegraphics{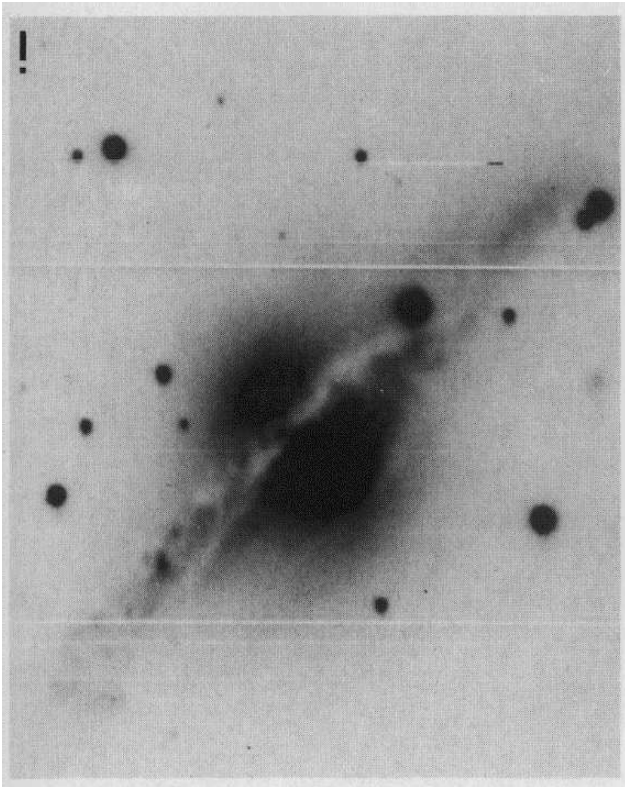}
\caption[]{NGC~4672 as it appears in Fig.~3i of the Polar-Ring
  Galaxy Catalog by Whitmore et al. (1990). The orientation of the 
  image is north to right and east bottom.}
\label{fig:n4672_whitmore}
\end{figure}

The galaxy is characterized by an intricate dust pattern ma\-de of
patches rather than lanes, which pervade the disk and cross the galaxy
near the center of the bulge  (Fig. \ref{fig:n4672_whitmore}).
NGC 4672 is a dust-rich system as one can also infer by a
comparison of its dust-to-blue luminosity ratio with the mean values
found by Bregman et al. (1992) for the early-type disk galaxies
studied by Roberts et al. (1991).  The dust pattern is so relevant in
determining the shape of NGC 4672 that Arp \& Madore (1987) included
it (with the name of AM~1243-412) in the section of the Catalogue of
Southern Peculiar Galaxies devoted to the galaxies with prominent
or unusual dust absorption.  Corwin et al. (1985) pointed out a
resemblance between the overall structure of NGC 4672 and that of
\object{NGC 5128} in their Southern Galaxy Catalogue. In spite of this
similarity with a dust-lane elliptical they classified NGC 4672 a
S0/a: pec sp galaxy since they recognized in the disk the presence of
spiral arms.  Recently, NGC 4672 has been included by de Grijs
(1998) in a large sample of edge-on disk galaxies which have been
observed in different optical and near-infrared passbands to study the
global properties of galactic disks.

\begin{figure}[ht]
\vspace*{9cm}
\includegraphics{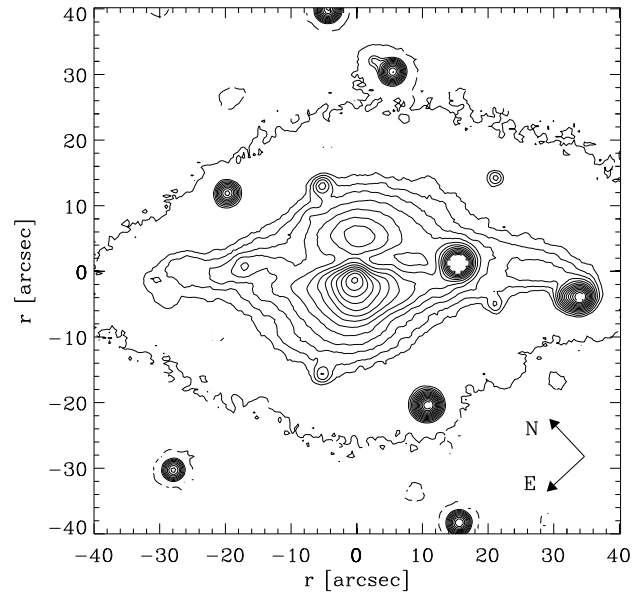}
\caption[]{$R-$band isophotes of NGC 4672 (boxcar smoothed over $3\times3$
  pixels). The inner isophotes are given in steps of 0.25 (uncalibrated) mag
  between -3.5 and -0.5 mag (see Fig.~\ref{fig:n4672_bul}), 
  while the outermost isophote corresponds to a level of 1.0 mag.} 
\label{fig:n4672_cont}
\end{figure}

Like in the case of NGC 4698 (Bertola et al. 1999), the bulge of
NGC 4672 appears elongated in an orthogonal way with respect to the
disk at a simple visual inspection of galaxy images 
(Fig.~\ref{fig:n4672_whitmore}). This remarkable feature is confirmed
by the shape of the inner ($|r|\la10''$) $R-$band isophotes, as shown
in the isophotal map of NGC 4672 in Fig.~\ref{fig:n4672_cont}.

For this peculiarity NGC 4672 has been included by Whitmore et
al. (1990) in the Polar-Ring Galaxy Catalog as a possible 
candidate polar-ring galaxy (PRC C-42). They stated that even if long
exposures show a fairly normal Sa galaxy, shorter exposures reveal
that the bulge component of the galaxy is slightly elongated in a
direction perpendicular to the disk. They considered this feature as a
signature of the structure of a polar-ring galaxy, since the observed
bulge and disk components of NGC 4672 could be interpreted
respectively as an almost face-on disk and a nearly edge-on ring
due to the orientation of the galaxy.

\begin{table}[ht]
\caption{Optical and radio properties of NGC 4672}
\begin{center}
\begin{tabular}{ll}
\hline
\noalign{\smallskip} 
parameter & value \\
\noalign{\smallskip} 
\hline
\noalign{\smallskip} 
Name                                            &  NGC~4672 \\
Morphological type           & .S.0$\ast$P/$\;^{\rm a}$;.SAS1P/$\;^{\rm b}$\\ 
Heliocentric systemic velocity$\;^{\rm c}$      & $3275\pm20$ \kms\\
Distance$\;^{\rm c}$                            & 39.9 Mpc\\
Disk position angle$\;^{\rm c}$                 & $46\dg$ \\
Apparent isophotal diameters $\;^{\rm b}$    & $122\farcs5\times33\farcs7$ \\ 
Inclination$\;^{\rm c}$                         & $75\dg$  \\
Apparent $B$ magnitude$\;^{\rm d}$              & $13.37$ mag \\
Apparent corrected $B$ magnitude$\;^{\rm b}$    & $12.96$ mag \\
Absolute corrected $B$ magnitude$\;^{\rm c}$    & $-20.04$ mag \\
Total $B$ luminosity $L_B^0$                    & $1.61\cdot10^{10}$ \lbsun \\
\hi\ linewidth at $20\%$ of the peak$\;^{\rm e}$& $403$ \kms \\
\hi\ linewidth at $50\%$ of the peak$\;^{\rm e}$& $356$ \kms \\
Mass of neutral hydrogen $M_{\rm HI}$$\;^{\rm f}$ & $2.88\cdot10^9$ \msun \\
$M_{\rm HI}/L_B^0$                                & $0.18$ \\
Mass of cool dust $M_{\rm d}$$\;^{\rm g}$       & $4.17\cdot10^6$ \msun \\
$M_{\rm d}/L_B^0$                                 & $2.59\cdot10^{-4}$ \\
\noalign{\smallskip} 
\hline
\noalign{\smallskip} 
\noalign{\smallskip} 
\noalign{\smallskip} 
\end{tabular}
\begin{minipage}{8cm}
$^{\rm a}$ from Corwin et al. (1985).\\
$^{\rm b}$ from de Vaucouleurs et al. (1991).  
The apparent isophotal diameters are measured at
a surface brightness level of $\mu_B = 25$ $\rm mag\;arcsec^{-2}$.\\ 
$^{\rm c}$ from this paper. The distance is derived as $V_0/H_0$ with
$V_0$ the velocity relative to the centroid of the Local Group
obtained from the heliocentric systemic velocity as in Sandage \&
Tammann (1981) and $H_0 = 75$ \kms\ Mpc$^{-1}$. The inclination $i$ is
derived as $\cos ^{2} i = (q^2-q_0^2)/(1-q_0^2)$, where the observed
axial ratio corresponds to $q=0.28$ (de Vaucouleurs et al. 1991) and
an intrinsic flattening of $q_0 = 0.11$ has been assumed following
Guthrie (1992).\\
$^{\rm d}$ from de Grijs (1998).\\
$^{\rm e}$ from Aaronson et al. (1989).\\
$^{\rm f}$ from Richter et al. (1994) for the adopted distance.\\
$^{\rm g}$ derived following Young et al. (1989) from the 
IRAS flux densities of NGC 4672 at 60 and 100 $\mu$m (Moshir et al. 1990).\\
\end{minipage}
\end{center}
\label{tab:n4672_properties}
\end{table}

\begin{figure}[ht!]
\vspace*{11cm}
\includegraphics{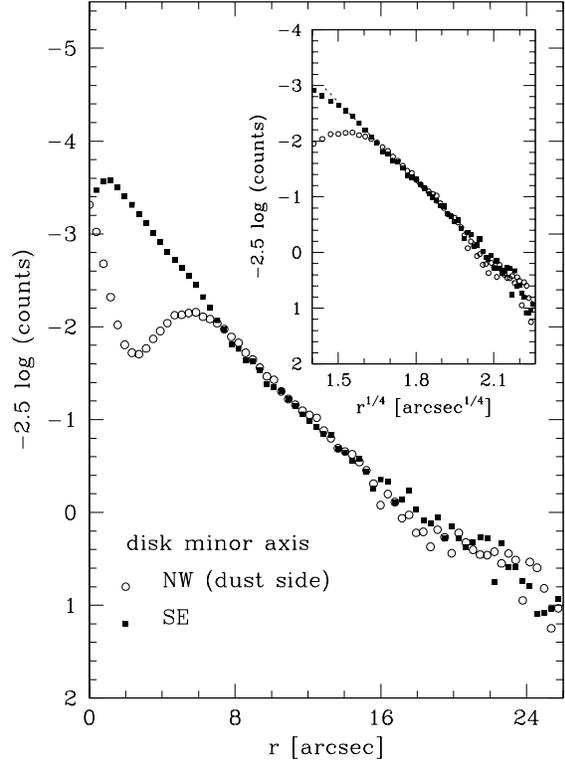}
\caption[]{Uncalibrated luminosity profile of NGC 4672 extracted along
the major axis of the bulge component ({\it i.e.\/}, the galaxy minor
axis at $\rm P.A.=134\dg$). {\it Filled squares\/} and {\it open
circles\/} represent the data obtained along the SE and the NW (dust
affected) sides, respectively. The folding has been performed to
obtain the best match of the luminosity profile on the two sides in
the outer, dust-free region.  In the inset the data are plotted
adopting a $r^{1/4}$-radial scale and the {\it dashed-line\/}
represents the best-fitting $r^{1/4}$ profile to the SE side data
($r_e = 9''$).}
\label{fig:n4672_bul}
\end{figure}

However there is a number of reasons to reject the polar-ring
classification and to consider the edge-on component of NGC 4672 a
disk rather than a polar ring, and the face-on component a bulge
rather than a low-inclined disk:

\begin{itemize}

\item[{\it (i)\/}] From the uncalibrated $R-$band acquisition image we
extracted the luminosity profile at $\rm P.A.=134\dg$ along the major
axis of the almost face-on component ({\it i.e.\/}, along the
minor axis of the edge-on component). It turns out that this profile
follows an $r^{1/4}$ law with an effective radius $r_e = 9''$
(Fig.~\ref{fig:n4672_bul}). This indicates that we are looking at a
stellar spheroid ({\it i.e.\/}, the bulge of an early-type disk galaxy
or an elliptical galaxy) rather than a face-on disk.

\item[{\it (ii)\/}] The edge-on component contributes a large fraction
of light to the total luminosity of the galaxy. In fact the
disk-to-total luminosity ratio we derived from de Grijs (1998) is
$D/T=0.75$ in the $B-$band, and $D/T=0.68$ in the
$I-$band.  The $B-$band $D/T$ is a typical value for an
intermediate-type spiral galaxy (Kent 1985; Simien \& de Vaucouleurs
1986).

\item[{\it (iii)\/}] The ionized-gas rotation curve measured along the
major axis of the edge-on component ($\rm P.A.=46\dg$) exhibits the
differential rotation (Fig.~\ref{fig:n4672_maj}) typical of a disk
component.  The absence of the characteristic linear rise of the gas
rotation velocity due to the so-called `rim of the wheel' effect ({\it
e.g.}, Vega Beltr\'an et al. 1997) is an indication that we are not
facing an edge-on ring.

\end{itemize}

Summarizing, NGC 4672 edge-on component can not be a ring-like
structure rotating around a low-inclined disk galaxy. Therefore the
possibility that NGC 4672 could be a polar-ring galaxy dressed up as a
spiral is ruled out. As a consequence, we assume as major-axis
position angle of NGC 4672 that of the disk component ($\rm
P.A.=46\dg$) instead of the value of $\rm P.A.=134\dg$ given by de
Vaucouleurs et al. (1991) and corresponding to the apparent major axis
of the bulge.

NGC~4672 was observed in the 21-cm neutral hydrogen line by
Aaronson et al. (1989) and by Richter et al. (1994).  The large \hi\
linewidth suggests that most of this gas is associated with the
edge-on disk.

\section{The kinematics of NGC 4672}
\label{sec:n4672_kinematics}

\subsection{Heliocentric systemic velocity}

\begin{figure}[ht]
\vspace*{8.5cm}
\includegraphics{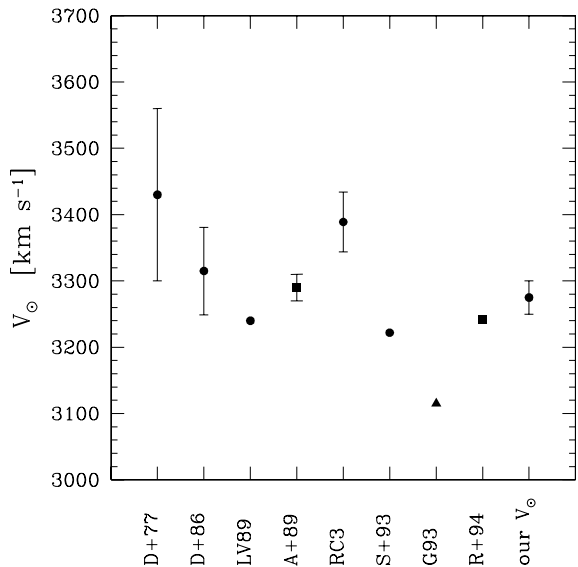}
\caption[]{Heliocentric systemic velocity of NGC 4672 derived from
  optical ({\it circles\/}), radio ({\it squares\/}) and optical and
  radio ({\it triangles\/}) measurements. D+77 = Dawe et al.
  (1977), D+86 = Dickens et al. (1986), LV89 =
  Lauberts \& Valentijn (1989), A+89 = Aaronson et al. (1989), RC3 = de
  Vaucouleurs (1991), S+93 = Schommer et al. (1993), G93 = 
  Garcia (1993), R+94 = Richter et al. (1994), our $V_\odot$ = this paper.} 
\label{fig:n4672_helio}
\end{figure}

We derived the heliocentric systemic velocity of NGC 4672 
$V_\odot = 3275 \pm 20$ \kms\ by fitting the major-axis rotation curves 
with a suitable odd function. 
In Fig.~\ref{fig:n4672_helio} we compare our heliocentric
systemic velocity with previous determinations based on optical and
radio measurements. It is in agreement wi\-thin the $3\sigma$
error with the values given by Dawe et al. (1977,
$V_\odot = 3430 \pm 130$ \kms), Dickens et al. (1986,
$V_\odot = 3315 \pm 66$ \kms), Lauberts \& Valentijn (1989, $V_\odot =
3240$ \kms), Aa\-ronson et al. (1989, $V_\odot = 3290 \pm 20$ \kms), de
Vaucouleurs et al.  (1991, $V_\odot = 3389 \pm 45$ \kms), Schommer et
al. (1993, $V_\odot = 3222$ \kms), Ri\-chter et al. (1994,
$V_\odot = 3242$ \kms). Only the systemic velocity given by Garcia
(1993, $V_\odot = 3115$ \kms) does not fall in this range.

\bigskip

\begin{figure*}[ht]
\vspace*{7.5cm}
\includegraphics{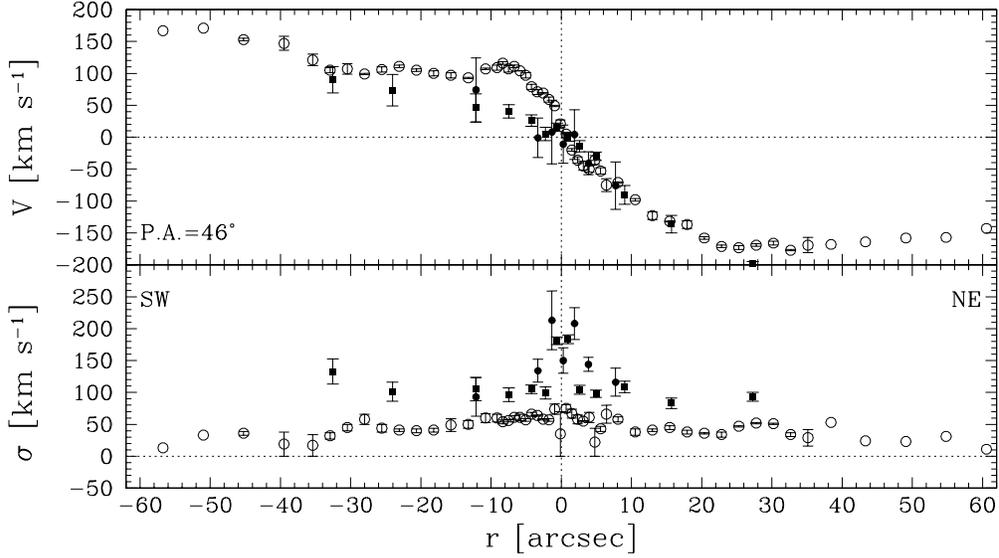}
\caption[]{The stellar ({\it filled symbols\/}) and ionized-gas ({\it
  open circles}) kinematics measured along the disk major axis ({\it
  i.e.\/}, the bulge minor axis at $\rm P.A.=46\dg$) of NGC 4672. The
  systemic velocity is $V_\odot = 3275 \pm 20$ \kms. All rotation
  velocities are plotted as observed without applying any inclination
  correction.  The {\it filled circles\/} and the {\it filled squares\/}
  represent data obtained in August 1998 and July 1999, respectively.}
\label{fig:n4672_maj}
\end{figure*}

The position-velocity curves and velocity dispersion profiles we
me\-asured for the stellar and gaseous components along the major and
minor axis of NGC 4672 are presented in Fig.~\ref{fig:n4672_maj} and
Fig.~\ref{fig:n4672_min}, respectively.

At each radius the plotted velocities $V$ of ionized gas and stars are
the observed ones after subtracting the value of the systemic
heliocentric velocity $V_\odot$ without correcting for the galaxy
inclination.

\subsection{Stellar kinematics}

The major-axis stellar kinematics is measured out to abo\-ut $30''$ on
each side of the nucleus (Fig.~\ref{fig:n4672_maj}). The velocity
curve is characterized by a central plateau showing no rotation for
$|r|\,\la\,3''$. At larger radii we observe an asymmetrical increasing
of the rotation velocity on both sides of the major axis. The
rotation velocity at the farthest measured point is about $90$ \kms\
and about $200$ \kms\ on the receding and approaching side,
respectively.  The velocity dispersion profile displays a central dip
of about $60$ \kms . In fact the velocity dispersion is $150\,\pm\,20$
\kms\ at $r\,\simeq\,0''$ and about $210$
\kms\ at $|r|\,\simeq1\,\farcs5$. Outwards it decreases reaching an
almost constant value of $100$ \kms\ for $|r|\,>\,6''$.

\begin{figure}[ht]
\vspace*{6.5cm}
\includegraphics{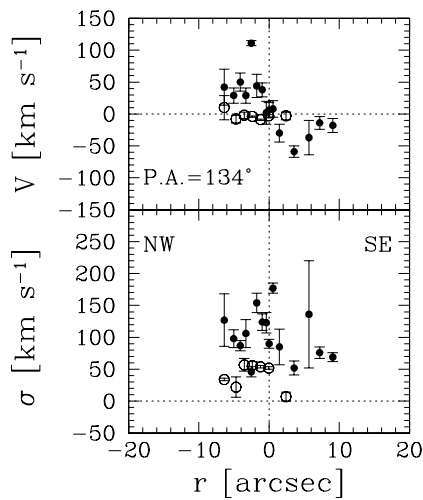}
\caption[]{The stellar ({\it filled circles\/}) and ionized-gas ({\it open circles})
  kinematics measured along the disk minor axis 
  ({\it i.e.\/}, the bulge major axis at $\rm P.A.=134\dg$) of NGC 4672.}
\label{fig:n4672_min}
\end{figure}

The minor-axis stellar kinematics extends to $-6''$ on the NW side and
to $+9''$ on the SE side, respectively (Fig.~\ref{fig:n4672_min}).
The velocity curve shows a steep gradient in the nucleus
($|r|\,\la\,3''$) rising to maximum rotation of about $85$ \kms\
($\Delta\,V_\star\,=\,170\,\pm\,18$ \kms). At larger radii it tends
to drop to a zero value on the SE side, while it remains almost
constant at about $40$ \kms\  on the NW side. Along the minor axis the
stellar velocity dispersion profile has a more uncertain behaviour. It
seems to fall from a central value of $154\,\pm\,15$ \kms\ to about
$80$ \kms\ further out.

\subsection{Ionized-gas kinematics}
 
The major-axis ionized gas kinematics is measured to ab\-out $60''$ from
the center on both sides of the galaxy
(Fig. \ref{fig:n4672_maj}). The gas velocity curve is radially
asymmetric.  On the SW side it exhibits a sharp gradient reaching
about $110$ \kms\ at $-7''$. Further out it shows bumps and wiggles
and the velocity oscillates between $93 \pm 3$ \kms\ and $111
\pm 3$ \kms\ in the radial range between $-7''$ and $-33''$. Then it
rises to about $170$ \kms\ at $-57''$.  Along the NE side the rotation
velocity increases almost linearly with radius from $0$ \kms\ to about
$170$ \kms\ between $0''$ and $+25''$, then it remains approximately
constant up to $+35''$. Outside it declines to about $140$ \kms\ at
$+61''$.  The gas velocity dispersion is lower than $70$ \kms\ at all radii.

Along the minor axis the measurements extend between $-6''$ and $+2''$
(Fig.~\ref{fig:n4672_min}).  No rotation is detected and the
corresponding velocity dispersion declines from about $60$ \kms\ to
$0$ \kms\ moving outwards from the center.

The fact that the same velocity dispersion corresponds to different
rotation velocities on both sides of the galaxy along its major
axis, shows that the ionized gas is an unreliable tracer of the
circular rotation in the observed radial range of NGC 4672.  Besides,
this kind of disturbed rotation curves are common in disk galaxies, as
shown in a recent study of the optical velocity curves of a
large number of Virgo S0's and spirals by Rubin et al.
(1999).

\bigskip

The analysis of the interplay between the ionized gas and stellar
kinematics along both the major and minor axis shows a decoupling
between the rotation of the gas and that of the inner ($|r|<5''$)
stellar component. Similarly, the stellar velocity curves
indicate that kinematical decoupling between the inner and outer
stellar components is also present as well.

\section{Discussion and conclusions}
\label{sec:n4672_conclusions}

In the previous sections we have pointed out that NGC 4672 is an
early-type disk galaxy (most probably an Sa spiral) with a prominent
bulge sticking out from the plane of the disk rather than a polar-ring
S0 galaxy. In addition to this remarkable orthogonal geometrical
decoupling between bulge and disk, the stellar velocity field of NGC
4672 is characterized by a central zero-velocity plateau along the
galaxy major axis and by a steep velocity gradient along the minor
axis. This means that in the inner regions of the bulge we are in the
presence of a stellar core, which is rotating perpendicularly to the
disk. We conclude that NGC 4672 has the same morphological ({\it
i.e.\/}, a bulge elongated perpendicularly to the disk) and
kinematical ({\it i.e.\/}, a stellar velocity gradient along the bulge
major axis) properties as those observed recently by Bertola et
al. (1999) in the Sa spiral NGC 4698.

\bigskip

As pointed out by Bertola \& Corsini (2000), the stellar rotation
curve measured along the bulge major axis (corresponding to the disk
minor axis) and the bulge minor axis of NGC~4672 and NGC 4698 
both show the same radial trends (characterized by an inner
velocity gradient and a zero-velocity plateau) which have be\-en
observed in the polar-ring elliptical AM 2020-504 along its major axis
(corresponding to the ring minor axis) and minor axis, respectively.

AM 2020-504 is considered the prototype of ellipticals with a polar
ring.  It is constituted by two distinct structures: a mostly gaseous
ring and a spheroidal E4 stellar component with their major axes
perpendicular each other.  It has been extensively studied by Whitmore
et al. (1987, 1990), who first revealed the presence of a velocity
gradient along the major axis of the spheroi\-dal component, and by
Arnaboldi et al. (1993a,b), who showed that the inner velocity
gradient of the stars is followed at larger radii by a velocity
decline to an almost constant zero value.  Recent observations
presented in Bertola \& Corsini (2000) are consistent with these
findings but also reveal that along the ring major axis the stars
exhibit a zero-velocity plateau followed by a rising velocity curve,
as in the cases of NGC 4698 and NGC 4672. Moreover no rotation of the
ionized gas is observed along the spheroid major axis, contrary to the
prediction of the warped model for the gaseous component by Arnaboldi
et al. (1993a).

\bigskip 

In spite of their morphological differences the early-type disk galaxy
NGC 4672, the Sa spiral NGC 4698 and the polar-ring elliptical AM
2020-504 share two common characteristics:

\begin{itemize}

\item[{\it (i)\/}] The major axis of the disk (or ring) component
forms an angle of $90^\circ$ with the major axis of
the bulge (or elliptical) component, at least at a visual inspection. 
This orthogonal geometrical decoupling is uncommon among spiral galaxies
(Bertola et al. 1991).

\item[{\it (ii)\/}] The stellar kinematics is characterized by
a central zero-velocity plateau followed by a rising velocity curve
along the disk (or ring) major axis, and by a steep velocity gradient
followed by an almost zero constant velocity along the disk (or ring)
minor axis. This indicates the presence of a kinematically isolated
core, which is rotating perpendicularly with respect to the disk (or
ring) component.

\end{itemize}

\bigskip

At this point we ask ourselves whether these galaxies also share
similar formation processes. For NGC 4698 Bertola et al. (1999)
suggested two alternative scenarios to explain the geometrical and
kinematical orthogonal decoupling between the inner region of the
bulge and the disk depending on the adopted (parametric or
non-parame\-tric) photometric decomposition.

If the parametric decomposition is adopted for NGC 4698, its
surface-brightness distribution is assumed to be the sum of an
$r^{1/4}$ bulge and an exponential thin disk. In this case the entire
bulge structure is found to be elongated perpendicularly to the major
axis of the disk.  The fact that the velocity field of the bulge is
characterized by zero velocity along the disk major axis and by a
velocity gradient along the disk minor axis suggests that the
intrinsic angular momenta of bulge and disk are perpendicular.  This
orthogonal decoupling of bulge and disk in NGC 4698 has been
interpreted  as the result of the formation of the disk component
at a later stage, due to the acquisition of material by a triaxial
pre-existing spheroid on the principal plane perpendicular to its
major axis.  For NGC 4698, according to the results of the
photometric decomposition the peculiar velocity curve observed along
the minor axis is representative of the rotation of its
bulge. This is also true for NGC 4672, since the bulge contribution
dominates the galaxy light along the minor axis in the region where we
measured the stellar kinematics.  However such a kind of rotation
curve (which is quickly rising to a maximum value and then fall
to about zero) is unrealistic for bulges, which are generally
believed to be isotropic rotators ({\it e.g.\/}, Kormendy \&
Illingworth 1982; Jarvis \& Freeman 1985a,b).

If, on the other hand, the non-parametric photometric
decomposition is adop\-ted for NGC 4698, the surface-brightness
distributions of bulge and disk do not depend on {\it a priori\/}
fitting laws but are assumed to have elliptical isophotes of constant
flattening. In this case the bulge of NGC 4698 should be round with
only the inner portion elongated perpendicularly with respect to the
disk major axis. For this reason the presence of an orthogonal
decoupled core has been invoked by Bertola et al. (1999) to explain
the observed photometric and kinematic  properties. This stellar
core should be photometrically decoupled from the disk and
kinematically decoupled with respect to both bulge and disk.  
With this in mind, it should be noticed that also the velocity fields
of NGC 4672 and AM 2020-504 show the same discontinuity between the
rotation of the inner and the outer regions as observed in NGC
4698. In elliptical galaxies such a discontinuity in the velocity
field is the signature of the presence of a kinematically-decoupled
core. Several formation scenarios have been proposed to explain the
origin of isolated cores, but the decoupling between the angular
momentum of a core and that of the host galaxy is generally
interpreted as the result of a second event (see Bertola \& Corsini
1999 for a review).
 
In NGC 4672 and AM 2020-504 the orthogonal geometrical decoupling
between bulge and the disk (or ring) components is directly visible
due to the high inclination of these galaxies.  This geometrical
decoupling  implies that such objects did not form in a single
event, as in the case of polar ring galaxies which are believed to be
the result of an accretion event (Steiman-Cameron \& Durisen 1982;
Schweizer et al. 1983; Sparke 1986) or a merger (Bekki
1998).  However it is hard to imagine that in NGC 4672 and AM 2020-504
the orthogonal disk component and the isolated core are the result
of {\it two\/} distinct and unrelated accretion phenomena
which occurred in different epochs.

Arnaboldi et al. (1993a,b) proposed a mechanism for the joint
formation of both the kine\-ma\-tically-decoupled core and the polar
ring of AM 2020-504.  Following Sparke (1986), they considered the
acquisition at a given angle of a large amo\-unt of external material
by an oblate elliptical galaxy.  Self-gravity acts on the gaseous
disk, which is initially forming at constant inclination, to transfer
angular momentum about the galactic pole from the outer material to
that inside.  In this way the outer gas moves towards nearly polar
orbits forming the observed polar ring, while the inner gas is pushed
away from the pole settling onto the equatorial plane of the host
galaxy, and subsequently turning into the stars, which now constitute
the isolated core.  It is interesting to observe that for $|r|>12''$
the kinematics of AM 2020-504 shows a non-rotating stellar body even
if the galaxy has been classified as an E4 elliptical and modeled as
an oblate spheroid (Arnaboldi et al.  1993a).  Anyway there are
a number of boxy E4 galaxies which show, slow or even no
rotation of the stellar body, like \object{NGC 1600} (Bender et al.
1994), \object{NGC 5322} (Scorza \& Bender 1995), and 
\object{NGC 5576} (Bender et al. 1994). In particular, 
the case of NGC~1600 has been discussed recently by Matthias \&
Gerhard (1999), who concluded that the kinematics of the inner
parts is consistent with a mostly radially anisotropic, axisymmetric
three-integral distribution function.

This scenario has the attractive property that it explains both the
orthogonal geometrical and kinematical decoupling between the
sphe\-roid and the ring components, as well as the kinematical
isolation of the core observed in AM 2020-504, as the byproduct of an
unique second event (represented by the ske\-wed accretion of large
amount of external material around an oblate spheroid) and its
following dynamical evolution.  Ber\-tola \& Corsini (2000) suggested
that the same acquisition me\-chanism produced also the
kine\-ma\-ti\-cal\-ly-decoupled cores in the center of the two disk
galaxies NGC 4698 and NGC 4672.  Therefore NGC 4672 and NGC 4698 are
interpreted as the result of the disk accretion in a polar plane of a
pre-existing oblate spheroid. Neither a velocity gradient along the disk
minor axis nor a geometrical decoupling are expected if the
acquisition process produces a disk settled on the equatorial plane of
the spheroidal component.

The ionized-gas component displays a different behavior in the inner
regions of NGC 4698 with respect to NGC 4672 and AM 2020-504.  In NGC
4698 the gas rotation closely matches that of the stars (see Fig.~2 in
Bertola \& Corsini 2000). The observed kinematics suggests that in the
center of NGC 4698 we are facing the ionized gas associated with the
kinematically decoupled stellar component giving rise to the velocity
gradient measured along the disk minor axis.  On the contrary, no
significant gas rotation is detected along the disk minor axis of NGC
4672 (Fig.~\ref{fig:n4672_min}) and AM 2020-504 (see Fig.~2 in
Bertola \& Corsini 2000), indicating that the inner ionized gas
resides in the galaxy disk and is not associated with the isolated
core.  This can be explained if in NGC 4672 and AM 2020-504 the
ionized gas settled onto the symmetry plane of the galaxy entirely
transformed into stars leaving behind a nearly edge-on gas-depleted
region, while in NGC 4698 this transformation is still in progress.

\bigskip

To identify candidates that may show the same features as
NGC 4672 and NGC 4698 one should look for disk galaxies with an
almost round bulge, whose isophotes appear to be elongated
perpendicularly to the galaxy major axis after the subtraction of the
disk surface brightness.  To a first visual inspection of the
photographic plates reproduced on the Carnegie Atlas of Galaxies
(Sandage \& Bedke 1994) the cases of \object{NGC 2911} (S0$_3$(2) or
S0 pec, Plate 49), \object{NGC 2968} (amorphous or S0$_3$ pec, Plate
49), \object{NGC 4448} (Sa (late), Plate 69) and \object{NGC 4933}
(S0$_3$ pec (tides), Plate 49) appear worthy of further
investigation.

Multi-band photometry could reveal color differences between the
stellar populations of the spheroid and possibly of the isolated core
(as done by Carollo et al. 1997a,b for kinematically decoupled
cores in ellipticals), allowing an estimate of their relative
age if combined with high-resolution spectroscopy. Moreover, the
latter will give us a detailed overview of the kinematics in the
center of these galaxies and the analysis of the line profiles could
unveil the dynamical nature of these cores ({\it e.g.\/}, the case of
the isolated core in the gE galaxy NGC 4365 studied by Surma \& Bender
1995).

\bigskip

Several mechanisms have been proposed for bulge formation (see Wyse et al.
1997 for a review) in which the bulge formed before
({\it e.g.\/} by hierarchical clustering merging), at the same time
({\it e.g.\/} in a monolithic collapse) or after the disk ({\it
e.g.\/} as a results of a secular evolution of the disk structure).
Furthermore, there is evidence that disks and bulges can experience
accretion events via infall of external material or satellite galaxies
(see Barnes \& Hernquist 1992 for a review of the different
processes). Up to now none of these pictures is able to reproduce
all the observed properties of disk galaxies along the Hubble
sequence, leading to the idea that several of the mechanisms outlined in
these different scenarios could have played a role in forming a galaxy
(Bouwens et al. 1999).

The early-type disk galaxies NGC 4672 and NGC 4698 represent striking
examples of galaxies which drastically cha\-nged morphology during their
history and for which we suggest an inside-out formation process with
the disk component accreted around a previously formed spheroid,
according to the hierarchical clustering merging paradigm (Baugh et al.
1996; Kauffman 1996).
  
\acknowledgements

We thank Magda Arnaboldi and Linda S\-parke for the useful discussion
about the case of AM 2020-504. We are indebted to
Bradley Whitmore for kindly proving us the image of NGC 4672 
we used in Fig. 1. Marc Sarzi acknowledges the
Max-Planck-Institut f\"ur Astronomie in Heidelberg for the hospitality
while this paper was in progress. This research has made use of the
Lyon-Meudon Extragalactic Database (LEDA) and of the NASA/IPAC
Extragalactic Database (NED).

\end{document}